\documentstyle[11pt,epsfig]{IOPART}
\def\eion{{(e~+~ion)}\ }
\newcommand{\be}{\begin{equation}}
\newcommand{\ee}{\end{equation}}
\begin{document}
\jl{2}
\title{Resolution and accuracy of resonances in R-matrix cross sections}

\author{Franck Delahaye, Sultana N. Nahar and Anil K Pradhan}
\address{Department of Astronomy,
Ohio State University \\ Columbus, OH 43210, USA}
\author{Hong Lin Zhang}
\address{Applied Physics Division, Los Alamos National Laboratory \\
Los Alamos, NM 87545, USA}

\begin{abstract}
 We investigate the effect of resonances in photoionization and
recombination cross sections computed using the R-matrix method. 
Photoionization and recombination rates derived from high-resolution
cross sections for oxygen ions
are compared with earlier works with less resolution and accuracy, such
as in the widely used Opacity Project data. We find significant
differences in photoionziation rates for O~II metastable states, 
averaged over Planck functions 
corresponding to ionizing radiation fields, with respect to the
intrinsic accuracy of the calculations {\it and} improved 
resolution. Furthermore, for highly
charged ions other physical effects are also important.
Recombination rate coefficients, averaged over a Maxwellian distribution,
are extremely sensitive to the position and 
resolution of near-threshold resonances, and radiation damping, 
in (e + O~VII) $\longleftrightarrow$ O~VI + h$\nu$. Surprisingly
however, the effect on the monochromatic and the mean
Rosseland and Planck bound-free opacities is relatively small, but may be
potentially significant.
\end{abstract}

\section{Introduction}
 A precise treatment of resonances has always been of prime concern in coupled
channel calculations for electron impact excitation, photoionization,
and recombination. Sophisticated methods have been employed to resolve
and/or fit resonances in order that cross sections and averaged rates,
the quantities of interest in applications, may be computed accurately.
One of the main reasons that the powerful R-matrix method (Burke \etal
1971, Hummer \etal 1993, Burke and Berrington 1993) has been very
successful is because it enables efficient calculations at a large
number of energies to affect such a resolution. Following a single
diagonalization of the R-matrix for each total electron-ion symmetry
LS$\pi$ or J$\pi$, cross sections may be obtained at any number of
energies. The R-matrix method, and its relativistic variant the
Breit-Pauli R-matrix method (BPRM, Berrington \etal, 1995), 
have been employed in large-scale
calculations in the Opacity Project (OP, Seaton \etal 1994, {\it The
Opacity Project Team} 1995, 1996), and the Iron
Project (IP, Hummer \etal 1993) for radiative
and collisional cross sections for most
astrophysically abundant elements, nearly 200 atoms and ions. The OP/IP
work involves photoionization and electron scattering cross sections, as
well as transition probabilities.
The R-matrix method (and BPRM) has been further extended to a unified
treatment of electron-ion recombination including radiative and
di-electronic recombination (RR and DR) in an ab initio manner, which is also
self-consistent with R-matrix photoionization cross sections
(Nahar and Pradhan 1992, 1994a; Zhang and Pradhan 1997, Zhang \etal 1999,
Nahar and Pradhan 2003a).
A crucial aspect of the unified method for \eion recombination is the
high resolution imperative for the derivation of accurate rate coefficients.
However, some physical effects are also very important in determining 
the accuracy of computed parameters as we demonstrate in this {\it
paper}. The present work also addresses some of
the issues and results described by Ramirez and Bautista (2002).

\section{Results}
 The effect of resolution and accuracy of resonances depends on physical 
processes.
The height of photoionization resonances is theoretically unbounded, unlike
electron-ion excitation cross sections where it is limited by
the unitarity constraint on the scattering matrix. Therefore resonances
in photoionization cross sections may approach unphysical heights if all
relevant physical effects are not carefully considered. A complete
resolution of resonances in photoionization cross sections, particularly
for higher $n$ and $\ell$, is difficult to achieve. Neither can the
calculated resonances be individually verified by experiments, since
the measured cross sections are averages over the beam width and shape.
Nevertheless, R-matrix calculations for photoionization cross sections
have been experimentally verified, some computed in LS coupling prior to
experiments. For example, we note the extremely detailed comparison
between R-matrix cross sections and recent experiments on synchrotron
based light sources reported by the Reno/Berkeley group for O II
(Covington et al, 2001), the Aarhus group for C II (Kjeldsen et al.
1999, Nahar 2002), and the Paris group for O ions (Champeaux et al
2003, Nahar 2004). Near-threshold resonances need to be carefully resolved, 
often with relativistic fine structure. In the relativistic Fe~XVII
photorecombination calculations including 359 bound levels with $n <$
10, Zhang \etal (2001) explicitly delineate 2,985 resonances in a very small
near-threshold region 0 - 5eV.

\subsection{Photionization cross sections and rates}
 In Fig. 1 we compare high-resolution R-matrix photoionization cross 
sections for
the ground and two metastable states of O~II, $2s^22p^3(^4S^o,^2D^o,^2P^o)$, 
with the OP data from Topbase
(Cunto \etal 1993, http://vizier.u-strasbg.fr/OP.html, 
http://heasarc.gsfc.nasa.gov/topbase/home.html). The OP cross sections were 
the most accurate attempt at the time of computations to resolve resonances. 
In general, instead of a constant energy mesh, a quantum-defect mesh
was employed to enable similar resolution of
resonances along a Rydberg series as they get narrower with increasing
$n$. The quantum-defect mesh
ensures a pre-determined number of points (usually 100)
 to resolve each
range of effective quantum number ($\nu,\nu+1)$. Although this does not
guarantee complete resolution, and the OP cross sections
may still be under-resolved for certain applications such as
monochromatic spectral modeling, they are considerably better than
those with constant energy mesh. 

 The new results reported in this work are obtained as in the recent
calculations for
a number of oxygen ions, O~II~-~O~V, using a more extensive
eigenfunction expansion for the target ion than in the OP work; they
are in good detailed agreement with recent experimental measurements 
(Nahar 2003).
Extensive resonances structures are delineated, especially in the 
near-threshold region. As Fig. 1 shows, resonances are more completely
resolved in the present calculations, including those along a Rydberg
series converging on to excited thresholds. Moreover, one important point 
is that the resonances positions are at somewhat different energies 
compared to the corresponding OP data. 

Photoionization rates $ \Gamma_{PI}$ are computed using both the present and the
OP cross sections, assuming ionizing radiaton fields from a blackbody
source $(B_{\nu} (T))$ at radiation temperatures T represented by a Planck 
function.

\be
\Gamma_{PI}=W\int \frac{4\pi B_{\nu}(T)\sigma_{PI}(\nu)}{h\nu}d\nu 
\ee
\be
 B_{\nu}(T)=(2h\nu ^3/c^2)[exp(h\nu/k_BT)-1]^{-1}
\ee
where W is a dilution factor (set to 1 for the present study) and 
$\sigma_{PI}$ is the photoionisation cross section. 

We can see in Fig.1, the ground state cross sections agree well. The position 
of the resonances are identical. Their heights differ but, as shown below,
it should not affect the rates significantly because the area under these peaks
is not significant since the resonances are narrow. For the two  metastable
states a shift in the position of the resonances can be seen.
The difference in resolution also modifies
the computed shape of the resonances, especialy for the first ones. 
These two factors, 
resolution and accuracy of the position of the resonances, results in
significant deviations in photoionisation rates.  

In Fig.2 we present the percentage difference in the rates, obtained using OP 
cross sections and the present data,
for the three lower states of O~II at radiation temperatures ranging 
from $10^3 - 10^5$ K. While there is little differences for the 
ground state $^4S^o$,
because the cross sections agree very well in the position of the 
resonances as well as in the resolution between the two set of data,
the differences for the two metastable states are quite signficant, 
up to 30\% for $^2P^o$, and up to 55\% for the $^2D^o$. 
In order to ascertain whether the differences are due to resolution, or 
accuracy as reflected in the different positions of resonances, another 
set of rates are obtained on shifting the resonance positions by a slight
amount to match in the two sets of data. This reduces the differences
by 10\%, with new maximum differences of 20\% and 45\%. Resolution 
also plays a role in the discrepancy. For the metastable states 
($^2P^o,\ ^2D^o $), the first resonance is much broader in the OP data. 
The areas under this first peak are 1.8 and 1.9 times higher
respectively for the OP cross sections compared to the present set. This
affects the low temperature regime more severely since the first
resonance is crucial in convolving with the Planck function, although
less so than with a Maxwellian, considered next.  

\subsection{Recombination rates}
 A similar situation exists for the highly charged ion Li-like
 O~VI. In previous calculations it was shown that the large KLL
resonances were severly under-resolved in previous works, and a very
high resolution is required in order to obtain converged values for the
resonance oscillator strengths. The peak values of resonances rise more
that 4 orders of magnitude above the background cross sections. 
Moreover, since the core ion is He-like,
with a large radiative decay rate for the dominant core $1s2p (^1P^o_1) 
\rightarrow 1s^2 (^1S_0)$ transition, radiation damping is important
(Zhang \etal 1999). This is particularly so in the calculation of
unified photo-recombination rate coefficients for (e + O~VII)
$\longleftrightarrow$ O~VI + h$\nu$ (Nahar and Pradhan 2003b). The
unified scheme to obtain total recombination rates,
including RR and DR, entails: (A) highly resolved calculations of
photoionization cross sections of a large number of bound levels,
typically several hundred levels with effective quantum number $\nu \leq
10$, and (B) DR calculations using an adaptation of the Bell and Seaton
theory (1985, Nahar and Pradhan 1994a) for $10 < \nu \leq \infty$. In
addition, for any given ion self-consistency is assured in an ab initio
manner between total photoionization and recombination since the same
wavefunction expansion is employed for both processes (see Nahar
2003, and references therein). Another advantage of the method is that
level-specific cross sections and rates for all levels with $\nu \leq
10$ are also obtained. 

 Total and level-specific rate coefficients for recombination to O~VI and 
O~VII have been reported by Nahar and Pradhan (2003b).
 In a recent work, Ramirez and Bautista (2002) calculated photoionization
cross sections for the ground state of highly charged ions such as 
O~VI and Fe~XXIV using a method to locate resonances for
high resolution (Quigley and Berrington 1996), but
neglecting effects such as fine structure and radiation damping. Therefore
their photo-recombination rate coefficients are in error by large
amounts, as seen in Fig. 3; there is over an order of magnitude
discrepancy relative to Nahar and Pradhan (2003b).  
Thus in this case it is seen that both
resolution and accuracy, as reflected in the neglected physical effects,
are important. In their work Ramirez and Bautista (2002)
discuss the earlier photoionization and \eion recombination calculations
for Fe~XVII by Pradhan \etal (2001); however their uncertainly estimates
of ``25-50\%" for ``integrated resonance contributions" are not derived from
direct computations, and are much higher than those computed by Zhang \etal
(2001). 

\subsection{Opacity}
 Finally, we investigate these effects on a larger scale, in the
total opacity of O~II. Fig. 4 shows the monochromatic bound-free (bf) 
opacity due to O~II (in atomic units)
at a representative stellar temperature of $2 \times 10^4$ K and
electron density of $10^{17}$ cm$^{-3}$, computed as described in Seaton
\etal (1994).

\be
 \kappa^{bf}(\nu)=\sum_i N_i \sigma^{bf}_i(\nu)[1-e^{-\frac{h\nu}{k_BT}}] 
\ee

The top panel shows results using
the present high-resolution photoionization data for 329 bound states of O~II,
compared with results from OP data. At this temperature and density,
197 bound states of O~II contribute to the bf-opacity in the present
calculations, and 169 in the OP data. The difference in number of bound
states is inconsequential; the additional states in the present work are
mostly highly excited ones. The monochromatic opacity is sampled at $10^5$
points in both calculations.
As Fig. 4 shows, there is overall very good
agreement between the two sets, despite significantly better resolution
and accuracy of the present calculations. This is expected since strong
features dominate the opacity spectrum, and are found in both
sets of calculations. The present Rosseland and Planck mean bf-opacities

\be
 \frac{1}{\kappa^{bf}_R}=\int \frac{1}{\kappa^{bf}_\nu}F_R(\nu)d\nu 
\ee
\be
 \kappa^{bf}_P=\int \kappa^{bf}_\nu F_P(\nu)d\nu 
\ee

where
\be
 F_R(\nu)=\frac{15}{4\pi^4}(\frac{h\nu}{k_BT})^4\frac{e^{-\frac{h\nu}{k_BT}}}{(1-e^{-\frac{h\nu}{k_BT}})^2}
\ee
\be 
 F_P(\nu)=\frac{15}{4\pi^5}(\frac{h\nu}{k_BT})^3\frac{e^{-\frac{h\nu}{k_BT}}}{(1-e^{-\frac{h\nu}{k_BT}})}
\ee

are 0.159 and 2.12 respectively, compared to 0.146 and 2.42 from the OP data, 
i.e. the Rosseland means are 8\% higher, and the Planck means are 14\% lower
than OP (electron scattering opacity is included to provide a
background, and is the same in both calculations). 
In the present opacity calculations the resonances are better resolved, and 
in many cases narrower compared to
the OP data, which results in a smaller Planck mean; conversely, since
the minima in resonances are better sampled, the Rosseland mean is
slightly higher.
However, the total
Rosseland mean opacities differ only by 2.6\%,
if we include the same (much larger) bound-bound contribution from lines
in the two calculations --- 0.345 from the present vs. 0.336 from the OP
data.

\section{Conclusion}
In conclusion we note that:

 1. For photoionization and NLTE modeling applications, it may be necessary 
to compute photoionization cross sections with
higher accuracy and resolution than in the Opacity Project data.
Accuracy of
resonance positions is crucial in the determination of \eion and
photoionization rates. A number of self-consistent
computations are in progress as part of a program to calculate 
photoionization and unified \eion reocmbination (RR plus
DR) cross sections and rates (Nahar and Pradhan 2003a).

 2. Although the monochromatic and mean
opacities are not measurably different using the high-resolution data
for oxygen ions, as opposed to the Opacity Project data, these
quantities
may differ considerably for more complex atomic systems such as the low
ionization stages of iron. In a previous work (Nahar and Pradhan 1994b),
we showed that the Rosseland mean opacities using new improved data for
Fe~II differed by up to 50\% from OP values. There is increasing evidence 
that there is "missing UV opacity", possibly due to Fe~I, 
that is critical to the determination of 
abundances in the Sun and other objects (e.g. Asplund 2003).  
New atomic and opacities calculations for iron ions are needed with
higher accuracy and resolution.

\pacno{34.80.Kw}
\maketitle

\submitted

\ack

This work was partially supported by the US National Science Foundation
and NASA. The computational work was carried out on the Cray SV1
at the Ohio Supercomputer Center in Columbus, Ohio.

\section*{References}

\def\apj{{\it Astrophys. J.}\ }
\def\apjs{{\it Astrophys. J. Supp. Ser.}\ }
\def\apjl{{\it Astrophys. J. (Letters)}\ }
\def\aj{{\it Astron. J.}\ }
\def\pasp{{\it Pub. Astron. Soc. Pacific}\ }
\def\mn{{\it Mon. Not. R. astr. Soc.}\ }
\def\aa{{\it Astron. Astrophys.}\ }
\def\aasup{{\it Astron. Astrophys. Suppl.}\ }
\def\baas{{\it Bull. Amer. Astron. Soc.}\ }
\def\jqsrt{{\it J. Quant. Spectrosc. Radiat. Transfer}\ }
\def\jpb{{\it Journal Of Physics B}\ }
\def\pra{{\it Physical Review A}\ }
\def\prl{{\it Physical Review Letters}\ }
\def\adndt{{\it At. Data Nucl. Data Tables}\ }
\def\cpc{{\it Comput.Phys. Commun.}\ }

\begin{harvard}

\item[] Asplund 2003 astro-ph 0312291 (\aa in press)

\item[] Bell R H and Seaton M J 1985 \jpb {\bf 18} 1589

\item[] Berrington K A, Eissner W, Norrington P H 1995 \cpc {\bf 92} 290

\item[] Burke P G, Hibbert A and Robb W D 1971
\jpb {\bf 4} 153

\item[] Burke P G and Berrington K A, {\it Atomic and Molecular
Processes: an R-matrix Approach}, IOP Publishing, Bristol, 1993

\item[] Champeaux J -P, Bizau J -M, Cubaynes D, Blancard C, Nahar S N,
Hitz D, Bruncau J, and Wuilleumier F J  2003 \apjs {\bf 148} 583 

%\item[] Chen G X and Pradhan A K 2002 \prl {\bf 89} 013202
%\item[] Chen G X, Pradhan A K and Eissner W 2003 \jpb (in press)

\item[] Covington A M, Aguilar A, Covington I R, Gharailbeh M,
Shirley C A, Phaneuf R A, Alvarez T, Cisneros C, Hinojosa H,
Bozek J D, Dominguez I, Sant'Ama M M, Schlachter A S, Berrah N,
Nahar S N, McLaughlin B M 2001, Phys. Rev. Lett. {\bf 87}, 243002-1

\item[] Cunto W., Mendoza C., Ochsenbein F., Zeippen C.J. 1993, \aa 
{\bf 275} L5 

\item[] Hummer D G, Berrington K A, Eissner W, Pradhan A K, Saraph
H E and Tully J A 1993 {\it Astr. Ap.} {\bf 279} 298

\item[] Kjeldsen H, Folkmann F, Hensen J E, Knudsen H, Rasmussen M S, West
J B, and Andersen T 1999, Astrophys. J. {\bf 524}, L143

\item[] Nahar S N 2002 \pra {\bf 65} 5207

\item[] Nahar S N 2004 \pra {\bf 69} 042714 

\item[] Nahar S N and Pradhan A K 1992 \prl {\bf 68} 1488

\item[] Nahar S N and Pradhan A K 1994a \pra {\bf 49} 1816

\item[] Nahar S N and Pradhan A K 1994b \jpb {\bf 27}, 429

\item[] Nahar S N and Pradhan A K 2003a {\it Self-Consistent R-matrix
Approach To Photoionization And Unified Electron-Ion Recombination}, in
Radiation Processes In Physics and Chemistry, Elsevier (in press).

\item[] Nahar S N and Pradhan A K 2003b \apjs {\bf 149}, 239 

\item[] Pradhan A K 2000 \apj {\bf 545} L165

\item[] Pradhan A K, Nahar S N and Zhang H L 2001 \apj {\bf 549} L265

\item[] Quigley L and Berrington K A 1996 \jpb {\bf 29} 4529

\item[] Ramirez J M and Bautista M A 2002 \jpb {\bf 35}, 4139

%\item[] Rogers F J and Iglesias 1992 \apjs {\bf 79} 507
%
\item[] Seaton M J, Yu Y, Mihalas D and Pradhan A K 1994 \mn {\bf 266} 805

\item[] {\it The Opacity Project, Vols 1 \& 2}, 1995, 1996, compiled by
the Opacity Project Team (Institute of Physics Publishing, Bristol and
Philadelphia) --- {\bf OP}

\item[] Zhang H L and Pradhan A K 1997 \prl {\bf 78} 195

\item[] Zhang H L, Nahar S N and Pradhan A K 1999 \jpb {\bf 32} 1459

\item[] Zhang H L, Nahar S N and Pradhan A K 2001 \pra {\bf 64} 032719

\end{harvard}

%\Tables

\clearpage
\begin{figure}
\psfig{file=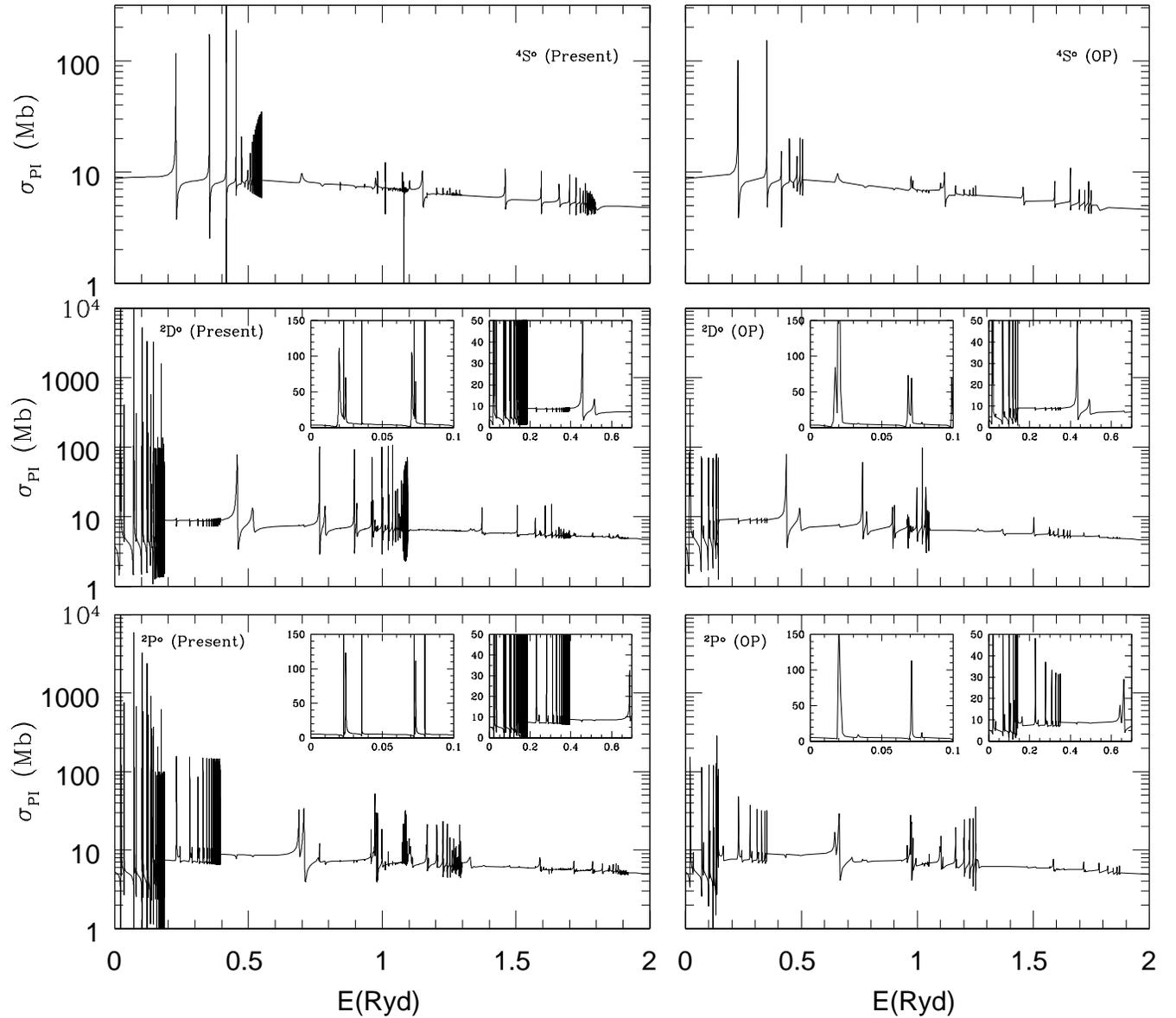,width=18cm}
\caption{Photo-ionization cross sections of the ground state 
$1s^22s^22p^3\ (^4S^o)$ and two meta-stable states 
$1s^22s^22p^3\ (^2P^o,\ ^2D^o)$ of O~II (from top to bottom). 
The present high-resolution data 
are on the left and the Opacity Project data from Topbase are on the right.
The insets show the shift in resonance positions and the effect of resolution 
in the near threshold region.}

\end{figure}

\clearpage
\begin{figure}
\psfig{file=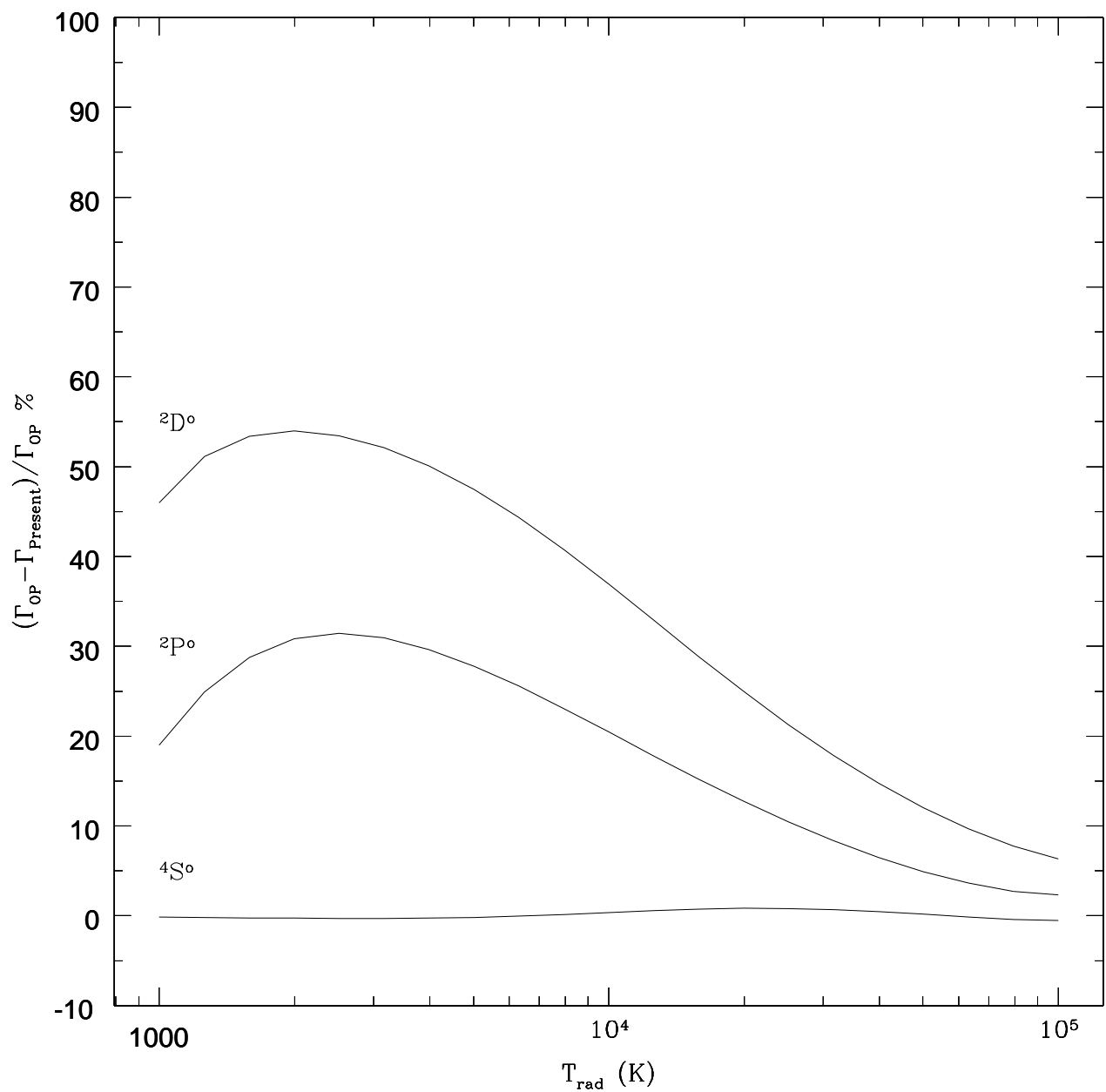,width=18cm}
\caption{Percentage differences in photoionization rates for the 
ground state $1s^22s^22p^3 (^4S^o)$ and metastable states 
$1s^22s^22p^3 (^2P^o, ^2D^o)$ of O~II, at radiation 
temperatures T$_{rad}$.}
\end{figure}

\clearpage
\begin{figure}
\psfig{file=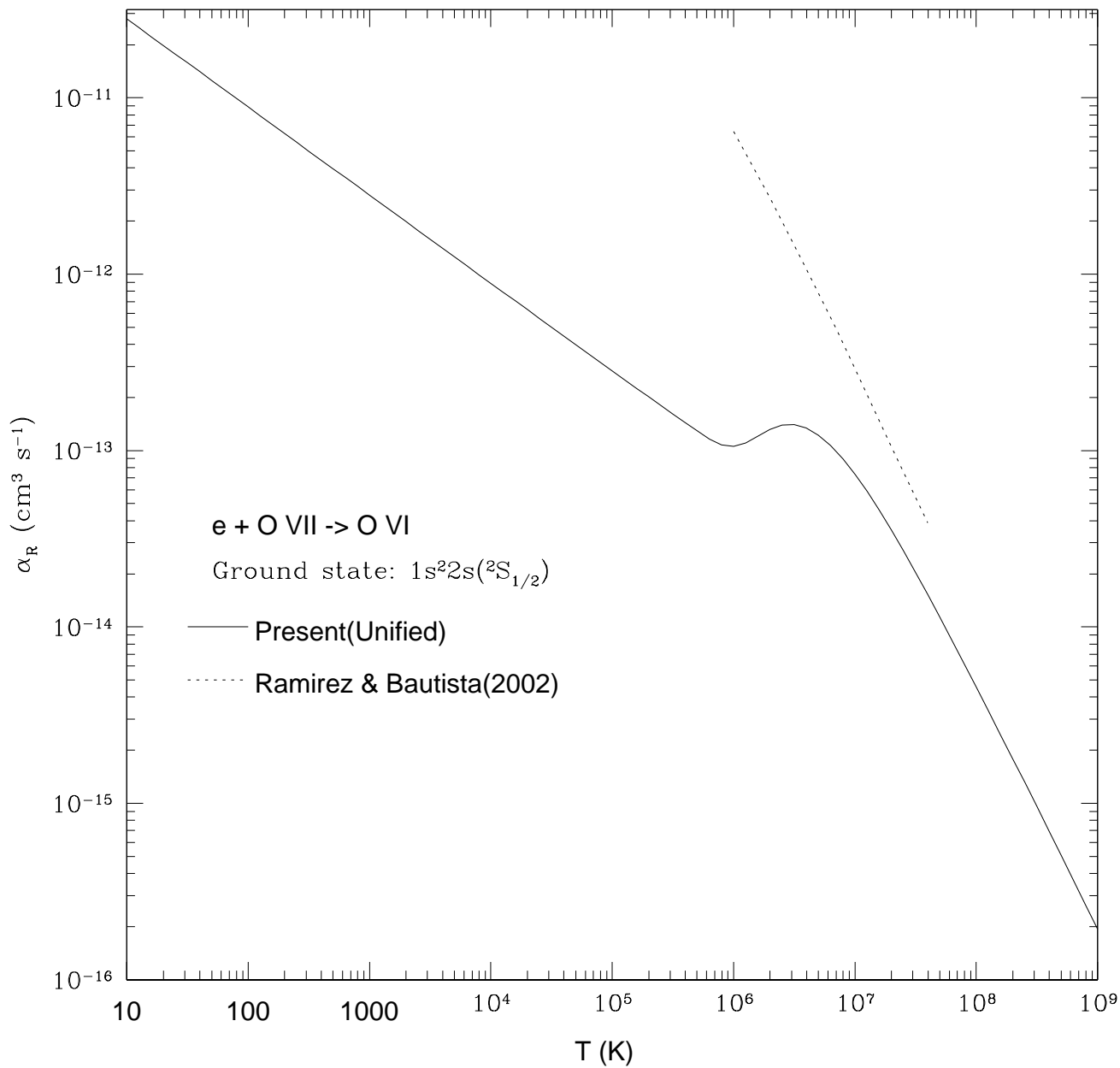,width=18cm}
\caption{Photo-recombination rate coefficient for the ground state of O~VI
($1s^22s (^2S_{1/2}$): solid line - using cross sections from Nahar and
Pradhan (2003), dashed line - Ramirez and Bautista (2002); the differences
are due to neglect of radiation damping in the latter calculations.}
\end{figure}

\clearpage
\begin{figure}
\psfig{file=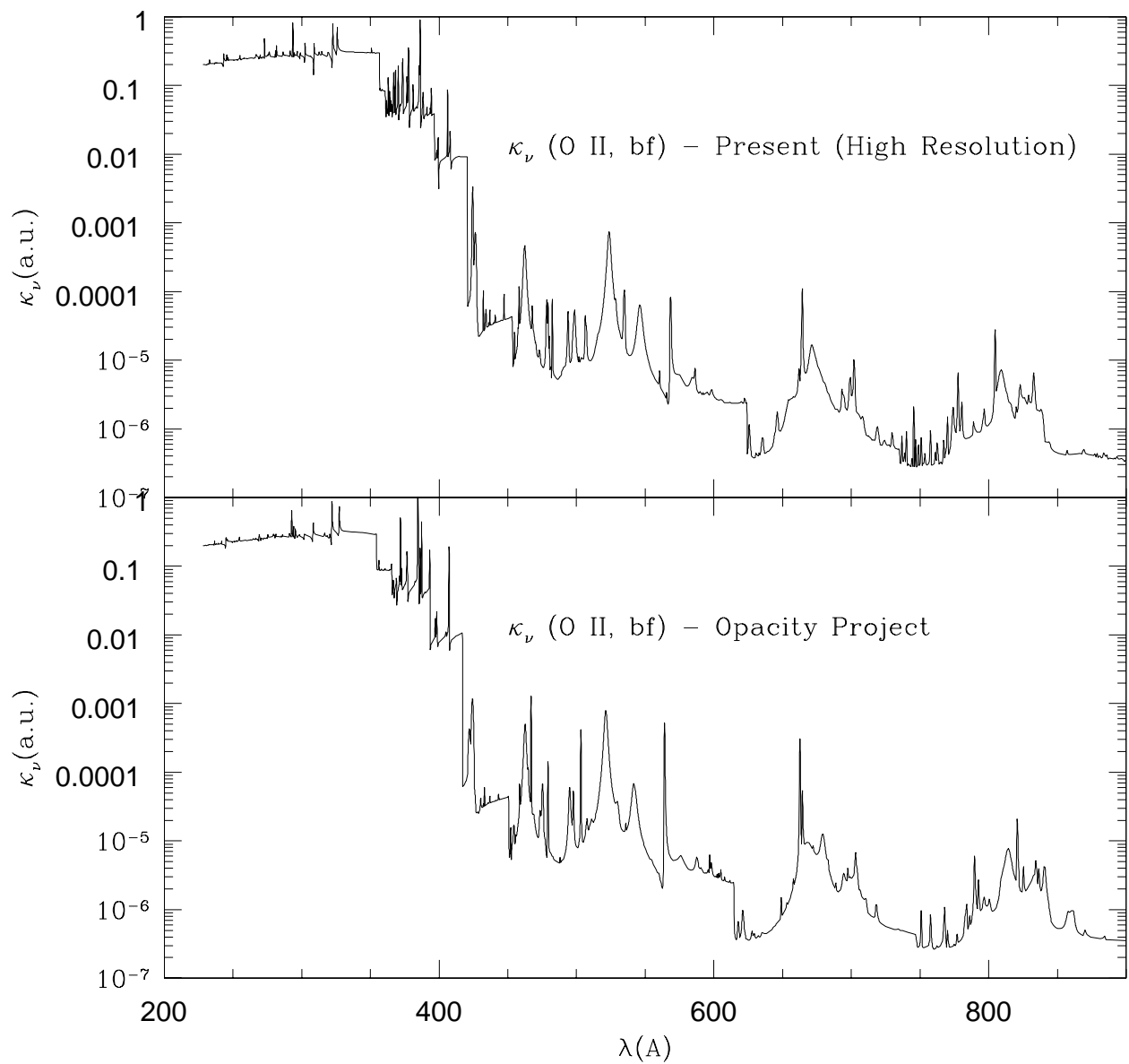,width=18cm}
\caption{Monochromatic bound-free opacity of O II using the present
high-resolution atomic data for photoionisation cross sections of 
329 bound states (upper panel), and using the Opacity Project data
for the same states (lower panel).}
\end{figure}
\clearpage

\end{document}